# A Machine Learning Framework for Quantum Cascade Laser Design

Andres Correa Hernandez and Claire F. Gmachl

*Department of Electrical and Computer Engineering, Princeton University, Princeton, NJ, 08545 USA*

A multi-layer perceptron neural network was used to predict the laser transition figure of merit, a measure of the laser threshold gain, of over 900 million Quantum Cascade Laser designs using only layer thicknesses and the applied electric field as inputs. Designs were generated by randomly altering the layer thicknesses of an initial 10-layer design. Validating the predictions with our 1D Schrödinger solver, the predicted values show 5% to 15% error for structures where a laser transition could occur, and 35% to 70% error for structures where there was no laser transition. The algorithm allowed (i) for the identification of high figure of merit structures, (ii) recognition of which layers should be altered to maximize the figure of merit at a given electric field, and (iii) increased the original design figure of merit of 94.7 to 141.2 eV ps $Å^2$, a 1.5-fold improvement and significant for QC lasers. The computational time for laser design data collection is greatly reduced from 32 hours for 27000 designs using our 1D Schrödinger solver on a virtual machine, to 8 hours for 907 million designs using the machine learning algorithm on a laptop computer.

## I. Introduction

Quantum Cascade Lasers (QCLs) are optoelectronic devices which mainly operate in the mid-infrared [1-3] and THz regimes [4, 5] of the electromagnetic spectrum. These lasers are useful for atmospheric trace chemical and particle detection [6, 7], low-visibility communication [8, 9], and label-free medical imaging [10, 11], among various other applications. A QCL design consists of alternating well and barrier material such that the bandstructure forms a multi-quantum well heterostructure. Electrons are pumped into the system and photons are emitted through intersubband transitions in the conduction band; between optical transitions the electrons continue traveling across the heterostructure through longitudinal optical (LO) phonon, and other scattering. Ideally, above laser threshold a photon is released for every electron and every period of the active region and injector that is present in the entire active core of the structure. The emission wavelength of the QCL can be tuned by changing specific parameters of the design such as layer thickness, applied electric field, material compositions, and material system. A combination of human intuition and computational analysis using solutions from a Schrödinger solver is a common approach to QCL design.

There have been several attempts at optimizing the large design parameter space of QCLs using various computational methods. Use of a genetic algorithm (GA) increased the wall-plug efficiency of a mid-infrared QCL by 7% [12], while another GA method was able to optimize a THz QCL transition frequency over a 2.9 THz range [13], among other applications [14-17]. Simulated annealing on a triple step quantum well design kept the transition energy around 50 meV [18], and also optimized superstructure gratings in QCLs to achieve non-equidistant frequencies [19]. Inverse spectral theory maximizes the gain in a quantum well laser [20] as well as optimizes the active region of a 12 μm QCL [21]. Machine learning (ML) for optimization of semiconductor devices has been applied to defect identification during the fabrication process [22-24] and to designing nanostructures in nanophotonics [25-27]. ML approaches for QCLs have so far looked at improving calculation time for modal gain [28], predicting resonant mode characteristics of QCLs in the THz regime [29], prediction of emission spectra of THz quantum cascade random lasers [30], predicting the threshold gain from higher-order modes given the refractive index profile of a QCL cavity [31], as well as the labeling of relevant wavefunctions [32]. In reference [33] an inverse network predicts the active region design of a QCL and then a forward network predicts metrics of interest from that design such as energy difference and LO lifetimes with low errors of 2-15 %, although entire QCL designs (active and injector regions) at different electric fields are not used for training the networks.

Here, we develop a framework for designing and optimizing QCL design by developing a ML tool around the Schrödinger solver. In particular, we develop a framework for optimizing a figure of merit (FoM), a measure for the laser threshold gain, for an initial QC design using ML. By specifying the number of layers, and layer thicknesses, along with the applied electric field, the algorithm predicts the FoM. Previous work started with collecting data for a QCL dataset to use in training an algorithm [34]. A laser transition code was built to compile QCL datasets by identifying the electronic state-pair transition in a QCL design with a high FoM in the bandstructure [35]. ML is then applied to various QCL datasets, optimizing the parameters of a multi-layer perceptron (MLP) neural network such as the activation functions and number of neurons in a hidden layer [36]. It was observed that ML worked well when predicting designs that were similar in structure to the starting design. This paper focuses on optimizing a FoM of a 10-layer structure by adding a random thickness to every layer in the -2 to +3 Å regime, an up to several 10% variation of layer thickness, and changing the electric field from 10 to 150 kV/cm. The total number of designs is predicted by ML and then selected designs with very high predicted FoM values are evaluated using our 1D Schrödinger solver. ML (i) suggests new QCL designs with high FoM, and (ii) recognizes which layers should be altered, and by how much, in order to maximize the FoM for a specific starting design.

## II. Methods

### A. Identifying Starting Design and QCL Parameters

A starting 10-layer structure was chosen (Fig. 1) consisting of three active region wells and two injection wells and was inspired by a successful 8.2 μm wavelength QCL [1]. The layer thickness sequence is **9**/57/**11**/54/**12**/45/**25**/34/**14**/33 (Å), where bold font indicates $Al_{0.48}In_{0.52}As$ barrier material and the normal text is $In_{0.53}Ga_{0.47}As$ well material. ErwinJr2, our research group's 1D Schrödinger solver [37], was used to calculate the eigenenergies and eigenfunctions of the multi-quantum well heterostructure at an applied electric field. ErwinJr2 also calculates the FoM, gain, and energy difference between any two eigenstates, among various other quantities. The FoM for this framework is calculated as

$$\text{FoM} = E_{ul}\tau_u \left(1 - \frac{\tau_l}{\tau_{ul}}\right) z_{ul}^2$$

where $E_{ul}$ is the energy difference between the upper and lower electronic subbands, $\tau_u$ and $\tau_l$ are the upper and lower scattering lifetimes, $\tau_{ul}$ is the LO phonon scattering lifetime between the upper and lower states, and $z_{ul}$ is the dipole matrix element. Our FoM has units of [eV ps Å$^2$] and is used to assess the quality of a state-pair transition, with a higher FoM indicating a better laser transition, leading e.g. to a lower threshold.

One design is defined as one QCL layer sequence at one electric field. Each design has a unique set of eigenenergies and eigenfunctions that change when any layer thickness, or the electric field, is altered. A unique aspect of QCLs is that there is a favorable state-pair interaction that releases photons at the energy difference of that electronic intersubband transition through stimulated emission. This "laser transition" is repeated for however many periods of active region of the structure. A code was developed to automatically identify this laser transition for every design using ErwinJr2 and can be found in reference [35]. The code collects the energy difference, scattering times, FoM, and gain of a laser transition from any design by specifying the layer thicknesses and electric field as inputs. Further details about the laser transition code can be found in reference [34].

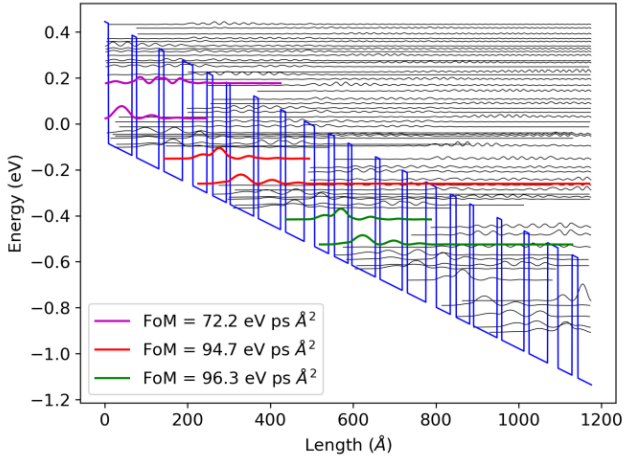

Fig. 1. Conduction band diagram of the initial 10-layer design at 90 kV/cm. The laser transition is highlighted in red and has a FoM of 94.7 eV ps Å$^2$.

Fig. 1 shows the bandstructure of our starting 10-layer design for four periods at 90 kV/cm. The laser transition code identifies the three state-pair electronic transitions with the highest FoM. The "laser transition" is identified as the state-pair energetically in the middle relative to the other state-pairs. The code marks it as red in graphs (if needed) and records it for a QCL dataset. For our starting design, the FoM of 94.7 eV ps Å$^2$ at 90 kV/cm was the highest for an electric field sweep of 10-150 kV/cm at 10 kV/cm increments. The energy difference is 108.9 meV or 11.4 μm emission wavelength.

### B. [-2, +3] Å Dataset

A dataset of 1800 structures was generated using the 10-layer starting structure, with an electric field range of 10-150 kV/cm applied to every structure in increments of 10 kV/cm. The 15-field iterations for 1800 structures gives a total of 27000 designs. The structures were generated by altering each of the 10 layers separately by a random thickness. This random thickness was an integer value and could vary from -2 to +3 Å, including 0 Å. For 10 layers and six integer random thicknesses, the total number of possible structures generated is $6^{10}$ or ~ 60.4 million. Including the electric field sweep, we have 15x$6^{10}$ or ~ 907 million total designs, which is our design space to be predicted by ML. The design space greatly increases as one layer, or one integer random thickness, is added, and why a 10-layer structure with [-2, +3] Å random tolerance range was initially selected when making the dataset. The dataset can be found at reference [38] and was generated with the Microsoft Azure cloud computing environment using a Linux Ubuntu virtual machine with 8 CPUs and 16 GB RAM. Other QCL datasets with different tolerance ranges, electric field sweeps, or based on different starting structures can be easily prepared in this same manner.

### C. Building the ML Algorithm

A multi-layer perceptron (MLP) neural network consisting of five layers is used to make the algorithm to predict QCL FoMs. The first layer of the neural network, the input layer, takes ten QCL layer thicknesses and the electric field of a design (11 inputs total). Next, three hidden layers follow, each consisting of a fully connected layer of 50 neurons, a normalization layer, and a rectifier linear unit (ReLU) activation layer. Lastly, the output layer has five quantities: FoM, gain, dipole matrix element, LO phonon scattering time, and energy difference of the selected laser transition, fully connected followed by a regression layer. This multi-output regression model uses the root-mean-square error (RMSE) to assess the accuracy of the FoM. The activation function used is the Adam optimizer with a 0.001 initial learning rate and trains for 150 epochs. The dataset is split up with 70% to train the network, 15% for validation, and 15% for testing. ML is done using the Deep Network Designer from MATLAB using a personal computer with 4 CPUs and 16 GB of RAM [39]. More details on building the MLP neural network and optimization of network parameters for different QCL datasets can be found in [36]. The trained ML algorithm can be found in reference [40].

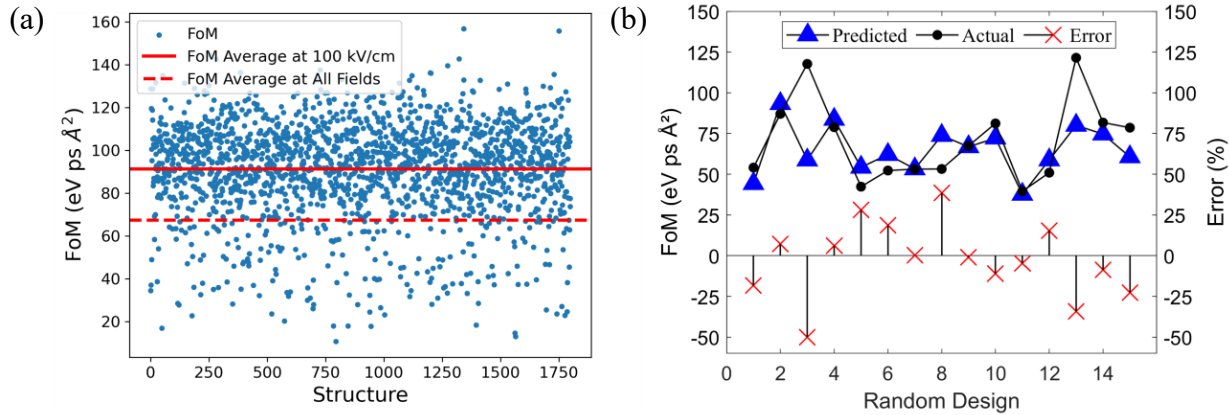

Fig. 2. (a) Plot of FoM for all 1800 structures at an applied field of 100 kV/cm. The solid line indicates the average FoM of 91.3 eV ps Å$^2$ at 100 kV/cm, while the dashed line is the FoM average of 67.3 eV ps Å$^2$ for all structures at every electric field. (b) FoM values for 15 designs randomly chosen from the test subset. The circles plot the actual ErwinJr2 FoM values and the triangles plot the predicted ML FoM values using the trained algorithm, with RMSE of 16.4 eV ps Å$^2$. The crosses are the error for each design.

### III. Results

#### A. [-2, +3] Å Dataset and ML Algorithm

27000 laser transition FoM values are collected in the [-2, +3] Å dataset [38] in approximately 32 hours using our Azure virtual machine. Fig. 2(a) is a FoM scatter plot of all 1800 structures at 100 kV/cm. Although there is only -2 to +3 Å variation between the layer changes, FoM values range anywhere from low 20 to high 150 eV ps Å$^2$, with the average FoM being 91.3 eV ps Å$^2$ (solid line). The average FoM for the entire dataset at all electric fields is 67.3 eV ps Å$^2$ and is plotted as a dashed line for reference.

The MLP neural network [40] was trained on the dataset in 412 seconds with a FoM RMSE of 16.4 eV ps Å$^2$. The FoM values for 15 designs of the test subset are plotted in Fig. 2(b) with the actual dataset values represented as circles, algorithm prediction depicted as triangles, and the error as crosses. The plot shows very good agreement between actual and predicted FoM values of the testing set. Designs 3, 8, and 13 had the highest error of 50.1%, 38.6%, and -34.2%, respectively. In these three cases the laser transition code did not collect a transition away from the edge boundary of the structure, thus giving numerical errors, inherent to the Schrödinger solver.

#### B. Design Subspace Visualization

The ML algorithm is used to predict the entire 10-layer [-2, +3] Å design space of 15x6$^{10}$ designs. Prediction of the entire design space is done under 8 hours by ML, 134000 times faster, and with less computational resources, than using the laser transition code. To visualize the data, the design space was separated into six subsets where the first layer of every design is kept constant. Thus, the total number of structures where +2 Å is always added to the first layer is $6^9$, and the total number of designs is 15x$6^9$ ~ 151 million.

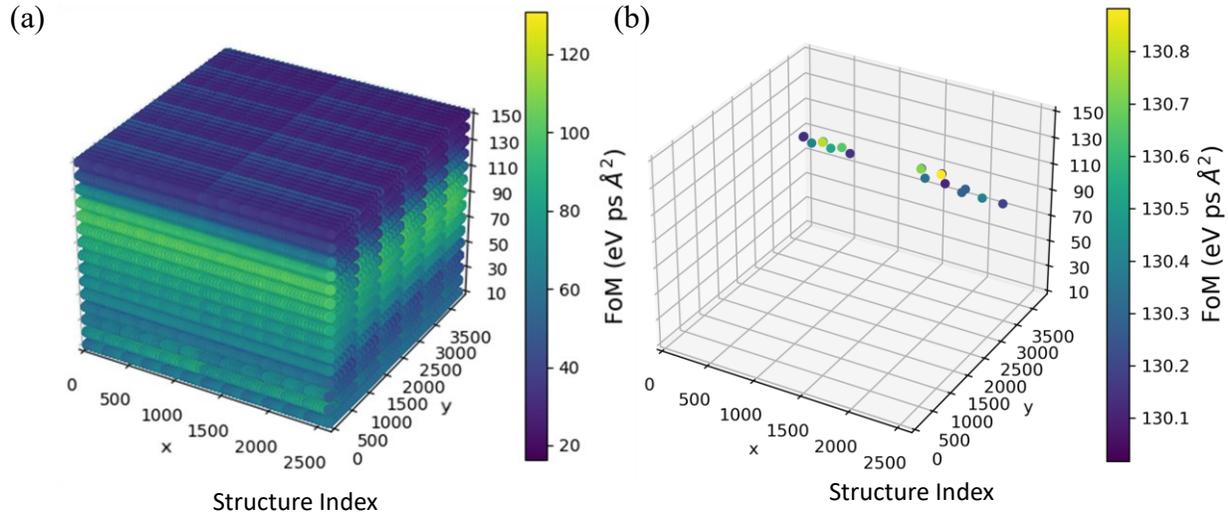

Fig. 3. 3D scatter plots of the ML FoM values for the [+2] Å subspace. The entire subspace is visualized in (a) with structures ordered sequentially in the x-y plane, the electric field in the z-axis, and the ML FoM expressed by color. (b) A FoM threshold filters out designs with FoMs lower than 130 eV ps Å$^2$, with the remaining 20 designs plotted.

The [+2] Å design subspace is visualized in Fig. 3(a) where every subspace structure is plotted in the x-y plane and every electric field iteration on the z-axis. The ML predicted FoM is visualized as the color of this 3D scatter plot. Every structure is organized in the x-y plane sequentially. The x-axis has length of 2x$6^4$ (2592) designs, and the y-axis has length 3x$6^4$ (3888) designs as this is the most linear and compact way of visualizing $6^9$ datapoints. The first design starts at the origin of the x-y plane, iterates through 2592 structures along the x-axis, varying the thickness of each layer, and then continues at structure 2593 on the second tick of the y-axis. Fig. 3(b) is the same 3D scatter plot but shows only designs where the FoM is higher than 130 eV ps Å$^2$. Applying a FoM threshold filter allows us to see where high performing structures are. From over 150 million structures plotted in Fig. 3(a), there are 20 with a FoM greater than 130 eV ps Å$^2$ shown in Fig. 3(b). The largest FoM designs are in the 90 and 100 kV/cm electric field range.

The highest FoM design is found at an applied electric field of 100 kV/cm. Fig. 4(a) shows the 2D scatter plot of all $6^9$ designs of the [+2] Å subspace at 100 kV/cm with the color representing the FoM. A FoM threshold of 129 eV ps Å$^2$ is applied and the 51 structures with FoM higher than this threshold are plotted as red dots on the 2D scatter plot. The highest ErwinJr2 FoM structure (denoted as S1) has an index (x, y) = (1506, 3241) on the scatter plot. Fig. 4(b) shows the bandstructure for this highest FoM design, S1, with the laser transition highlighted in red, ErwinJr2 FoM of 141.2 eV ps Å$^2$, and an error of -7.4% between ML and ErwinJr2 FoM.

### C. High FoM Structures Identified Using ML

Table I lists the top two FoM structures for all six subspaces. The layer addition, electric field, predicted ML FoM, actual ErwinJr2 FoM (EJ2), and error between the FoMs are also listed. Several layer thickness columns had identical values across all high FoM subspace designs indicating which layers are essential to best optimize the original 10-layer structure. These were layer 2 (L02) and layer 9 (L09) with +3 Å for all designs, as well as layer 5 (L05) and layer 10 (L10) with -2 Å for every design. The two highest FoM values for predicted ML designs happen in the [+3] Å subspace and the highest actual ErwinJr2 FoM values are in the [+2] Å subspace.

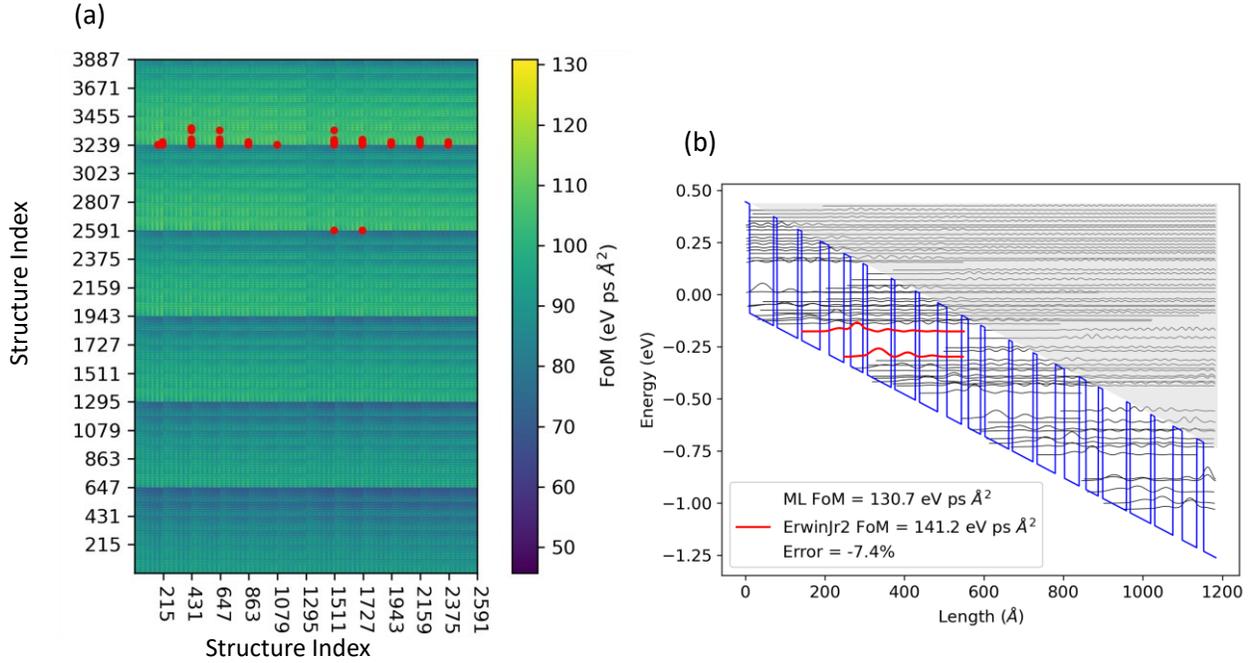

Fig. 4. The 2D scatter plot of the [+2] Å subspace is shown in (a) at 100 kV/cm with the color indicating the FoM values and 51 structures are shown with red dots at a FoM threshold above 129 eV ps Å$^2$. The bandstructure of S1 is shown in (b) with the highest FoM in the subspace of 141.2 eV ps Å$^2$, a ML FoM of 130.7 eV ps Å$^2$, and error -7.4%.

Design S1 from Fig. 4(b) is in the [+2] Å subspace and has the highest actual ErwinJr2 FoM of 141.2 eV ps Å$^2$ with a wavelength of 10.21 μm (121.4 meV) at 100 kV/cm. Design S1 is underlined and boldface in Table I. The original 10-layer QCL design has a FoM of 94.7 eV ps Å$^2$ at 90 kV/cm while the QCL design S1, identified by ML, increases the FoM 1.5-fold, a signification improvement for QCLs.

Table I. Highest FoM designs for the [-2, +3] Å Design Space

| L01 (Å) | L02 (Å) | L03 (Å) | L04 (Å) | L05 (Å) | L06 (Å) | L07 (Å) | L08 (Å) | L09 (Å) | L10 (Å) | E-Field (kV/cm) | FoM$_{ML}$ (eV ps Å$^2$) | FoM$_{EJ2}$ (eV ps Å$^2$) | Error (%) |
|---|---|---|---|---|---|---|---|---|---|---|---|---|---|
| -2 | 3 | -2 | -1 | -2 | -2 | -2 | 3 | 3 | -2 | 110 | 117.6 | 123 | -4.4 |
| -2 | 3 | -2 | 1 | -2 | -2 | -2 | 2 | 3 | -2 | 110 | 117.0 | 120.6 | -3.0 |
| -1 | 3 | -2 | 1 | -2 | -1 | 0 | 3 | 3 | -2 | 100 | 120.4 | 126.1 | -4.5 |
| -1 | 3 | -2 | 2 | -2 | -2 | 0 | 3 | 3 | -2 | 100 | 120.5 | 124.5 | -3.2 |
| 0 | 3 | -2 | 1 | -2 | -2 | 1 | 3 | 3 | -2 | 100 | 124.8 | 130.4 | -4.3 |
| 0 | 3 | -2 | 1 | -2 | -1 | 0 | 3 | 3 | -2 | 100 | 124.7 | 129.8 | -4.0 |
| 1 | 3 | -2 | -1 | -2 | 0 | -1 | 3 | 3 | -2 | 100 | 127.8 | 135.4 | -5.6 |
| 1 | 3 | -2 | -2 | -2 | 0 | 0 | 3 | 3 | -2 | 100 | 127.7 | 134.1 | -4.7 |
| **2** | **3** | **-2** | **-2** | **-2** | **1** | **-2** | **3** | **3** | **-2** | **100** | **130.7** | **141.2** | **-7.4** |
| 2 | 3 | -2 | -2 | -2 | 0 | -1 | 3 | 3 | -2 | 100 | 130.8 | 140.2 | -6.7 |
| 3 | 3 | -1 | -2 | -2 | 3 | -1 | 3 | 3 | -2 | 90 | 133.3 | 139.4 | -4.4 |
| 3 | 3 | 0 | -2 | -2 | 3 | -2 | 3 | 3 | -2 | 90 | 133.7 | 138.3 | -3.3 |

The second highest ErwinJr2 FoM design in the [+2] Å subspace (denoted S2), has the same operating field of 100 kV/cm, with a predicted ML FoM of 130.8 eV ps Å$^2$ and actual ErwinJr2 FoM of 140.2 eV ps Å$^2$ as seen in Table I. ErwinJr2 calculates the FoM for designs S1 and S2 that differ by only 1 eV ps Å$^2$, demonstrating the numerical sensitivity to layer thicknesses and boundary conditions of the 1D Schrödinger approach. Similarly, the predicted ML FoM values for these designs only vary slightly by 0.1 eV ps Å$^2$ showcasing the accuracy of the algorithm for identification of high FoM designs. There is only a 1 Å difference between structures S1 and S2 in layers six (L06) and seven (L07). The bandstructure of these QCL designs are very similar as well, thus indicating that these two layers, by themselves, do not play a major role in maximizing the FoM for the layer thickness tolerance range of [-2, +3] Å.

## Conclusion

A ML framework has been developed for optimizing the FoM for a 10-layer QCL design. A design space of $6^{10}$ structures with 15 electric field iterations, i.e. ~ 907 million unique designs, is created. The FoM for 27000 structures from this design space forms our initial dataset and is collected in 36 hours on a virtual machine. A MLP neural network splits the dataset into a training subset, a validation subset, and a testing subset, and afterwards is used to predict the entire design space of ~ 907 million designs in under 8 hours on a personal laptop computer. Different visualization techniques are used to plot this large data space by splitting the FoM values into six subspaces where the first layer thickness is constant. By comparing the layer thicknesses of high FoM structures in the subspaces with each other, ML (i) finds designs with high FoM, and (ii) identifies which layers are most important when maximizing the FoM for a starting design.

This ML framework is used to identify a new QCL design, S1, in the [+2] Å subspace, with a FoM 1.5 times larger than the original design. The design operates around the same electric field without adding or removing the number of layers, only changing the layer thicknesses by -2 to +3 Å and is identified faster than using the laser transition code or a 1D Schrödinger solver and human intuition. In the future, the framework will be used to expand the layer thickness design space and to identify new QCL design strategies.

## Acknowledgements

The authors gratefully acknowledge financial support from the Schmidt DataX Fund at Princeton University made possible through a major gift from the Schmidt Futures Foundation, the National Science Foundation under Grant No. DGE-2039656, and partially funded by the Center for Statistics and Machine Learning at Princeton University through the support of Microsoft. We thank Dr. Ming Lyu for the current version of ErwinJr2 [37] used in this paper.

## Author Declarations

### Conflict of Interest

The authors have no conflicts to disclose.

### Author Contributions

**Andres Correa Hernandez**: Conceptualization (equal); data curation (lead); formal analysis (lead); investigation (lead); methodology (equal); software (lead); validation (equal); visualization (lead); writing – original draft (lead); writing – review and editing (equal).

**Claire F. Gmachl**: Conceptualization (equal); formal analysis (supporting); funding acquisition

(lead); investigation (supporting); methodology (equal); supervision (lead); validation (equal); visualization (supporting); writing – original draft (supporting); writing – review and editing (equal).

**Data Availability**

The laser transition code is openly available in the Princeton Data Commons repository at https://doi.org/10.34770/z3r2-hg07, reference number [35].

The ErwinJr2 software is openly available to download on the GitHub repository at https://github.com/ErwinJr2/ErwinJr2, reference number [37]

The 10-layer [-2, +3] Å dataset is openly available in the Princeton Data Commons repository at https://doi.org/10.34770/r7nr-ee50, reference number [38]

The code to train the MLP neural network, as well as the algorithm used to predict the 900 million structures is available in the Princeton Data Commons repository at https://doi.org/10.34770/e034-4670, reference number [40].